\documentclass{iau}
\usepackage{graphicx}
\usepackage{aas_macros}

\title[The Formation of Globular Clusters] %% give here short title %%
{Are globular clusters the natural outcome of regular high-redshift star formation?}

\author[J.~M.~Diederik Kruijssen]   %% give here short author list %%
{J.~M.~Diederik Kruijssen}

\affiliation{Max-Planck Institut f\"{u}r Astrophysik, \\ Karl-Schwarzschild-Stra\ss e 1,
85748, Garching, Germany \\ email: {\tt kruijssen@mpa-garching.mpg.de}}

\pubyear{2015}
\volume{312}  %% insert here IAU Symposium No.
\pagerange{}
\jname{Star Clusters and Black Holes}
\editors{R. Spurzem, F. Liu, S. Li \& Y. Meiron}

\begin{document}

\maketitle

\begin{abstract}
We summarise some of the recent progress in understanding the formation and evolution of globular clusters (GCs) in the context of galaxy formation and evolution. It is discussed that an end-to-end model for GC formation and evolution should capture four different phases: (1) star and cluster formation in the high-pressure interstellar medium of high-redshift galaxies, (2) cluster disruption by tidal shocks in the gas-rich host galaxy disc, (3) cluster migration into the galaxy halo, and (4) the final evaporation-dominated evolution of GCs until the present day. Previous models have mainly focussed on phase 4. We present and discuss a simple model that includes each of these four steps -- its key difference with respect to previous work is the simultaneous addition of the high-redshift formation and early evolution of young GCs, as well as their migration into galaxy haloes. The new model provides an excellent match to the observed GC mass spectrum and specific frequency, as well as the relations of GCs to the host dark matter halo mass and supermassive black hole mass. These results show (1) that the properties of present-day GCs are reproduced by assuming that they are the natural outcome of regular high-redshift star formation (i.e.~they form according to same physical processes that govern massive cluster formation in the local Universe), and (2) that models only including GC evaporation strongly underestimate their integrated mass loss over a Hubble time.
\end{abstract}

\firstsection

\section{Introduction}
Approximately 50\% of all stars in the Universe formed at redshifts $z>1$, with a peak in the cosmic star formation history at $z=2$--$3$ (Madau \& Dickinson 2014). Understanding the cosmological assembly of stellar mass in galaxies therefore requires a census of the conditions under which star formation proceeded in high-redshift systems. Modern observational facilities enable detailed studies on the super-kpc scales of spatially-resolved galaxies at $z>1$ (Hodge et al.~2012; Tacconi et al.~2013), revealing gas pressures 3--4 orders of magnitude higher than in the local Universe (Swinbank et al.~2011; Kruijssen \& Longmore 2013). Despite these efforts, the cloud-scale ($\sim100~{\rm pc}$) physics of high-redshift star formation remain out of reach.

\begin{figure}
\center\includegraphics[width=12.5cm]{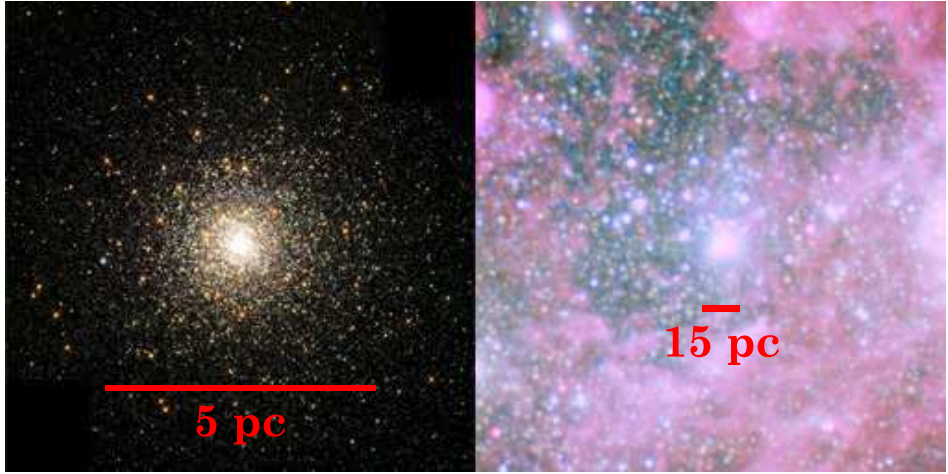}
\caption{{\it Left}: The Galactic GC M80, with a mass of $M\sim10^{5.6}~{\rm M}_\odot$ and a radius of $R\sim1.8~{\rm pc}$. Image credit: NASA and The Hubble Heritage Team (STScI/AURA). {\it Right}: The most massive young massive cluster in the nearby starburst NGC~1569, with a mass of $M\sim10^{6}~{\rm M}_\odot$ and a radius of $R\sim1.6~{\rm pc}$. Image credit: NASA, ESA, the Hubble Heritage Team (STScI/AURA), and A. Aloisi (STScI/ESA). The properties of both clusters are very similar, showing that special conditions exclusive to the early Universe are not necessary for the formation of GC-like clusters.} \label{fig:clusters}
\end{figure}

Globular clusters (GCs) have the potential to be used as tools for tracing extreme star formation physics in their high-redshift birth environments. However, the resulting insight is only relevant in the context of galaxy formation if GCs did not form through some `special' mode of star formation, but instead are the natural products of `regular' high-redshift star formation, i.e.~they result from the same physical processes that govern star formation in the local Universe. Given the extreme conditions seen in $z>1$ systems (e.g.~high gas pressures and densities), the observed old ages, high masses, high velocity dispersions, and low metallicities of GCs indeed seem to be consistent with the expected star formation products (see Figure~\ref{fig:clusters} and Elmegreen \& Efremov 1997; Shapiro et al.~2010; Kruijssen 2014). It is therefore reasonable to evaluate if GCs do indeed result from the dominant mode of star formation at $z>1$.

To establish whether GCs are indeed the natural outcome of regular, high-redshift star formation and eventually use them as fossils to trace the underlying physics, it is necessary to obtain an end-to-end description for GC formation and evolution. Only then, it is possible to correct for a Hubble time of GC evolution and `rewind' the current GC population to their birth sites. In this paper, we summarise our two most recent contributions to solving this problem.
\begin{itemize}
\item[(i)]
In Kruijssen (2014), we review GC formation in the context of galaxy formation and evolution.
\item[(ii)]
In Kruijssen (2015), we present a simple, end-to-end model for GC formation and evolution that combines our current knowledge of massive cluster formation, high-redshift star formation, present-day GC populations, and star cluster disruption.
\end{itemize}

Previous GC formation and evolution models typically rely on GC evaporation to explain the observed properties of present-day GC populations (most notably their mass spectrum, see e.g.~Fall \& Zhang 2001; Prieto \& Gnedin 2008; Kruijssen \& Portegies Zwart 2009; Li \& Gnedin 2014). However, these models often fail to reproduce other observables (e.g.~the specific frequency or the radial invariance within galaxies of the GC mass spectrum). The key new step made in the Kruijssen (2015) model is that it accounts for the `pre-processing' of the GC population by impulsive tidal shocks in their natal galaxy discs, which leads to their rapid tidal disruption well before evaporation becomes important (also see Elmegreen 2010), as well as the subsequent redistribution of GCs into galaxy haloes by hierarchical galaxy formation. By introducing this extra phase in the evolutionary history of the GC population, the model produces present-day GC populations that are consistent with those observed, e.g.~in terms of their mass spectrum, specific frequencies, and colours, as well as their relations to the host dark matter halo mass and supermassive black hole mass. The conclusion is that GCs are consistent with being the products of regular high-redshift star formation.

\section{Ingredients for an end-to-end GC formation and evolution model}
An end-to-end model for GC formation and evolution should describe the formation of GCs within their high-redshift natal environment, their early evolution within the host galaxy disc, their migration into the galaxy haloes, and their subsequent evolution until the present day. Here, we summarise the physics covered by the Kruijssen (2015) model.

\subsection{Cluster formation}
Studies of stellar cluster formation in the local Universe show that clusters form through the hierarchical collapse of density peaks in the interstellar medium (ISM) (Efremov \& Elmegreen 1998; Longmore et al.~2014; Rathborne et al.~2015). Whether the resulting stellar system becomes a gravitationally bound stellar cluster or an unbound association depends on the star formation efficiency (SFE) -- at low SFEs, the removal of the residual gas by feedback unbinds the stellar system (e.g.~Hills 1980; Lada et al.~1984; Goodwin \& Bastian 2006). Because the SFE increases with the number of free-fall times completed before residual gas removal, the highest density peaks end up being gas-poor and remain gravitationally bound (Kruijssen et al.~2012a). By integrating the resulting bound fraction of young stars over the density spectrum of the ISM, one can formulate a model to predict the fraction of all star formation in a galaxy that results in bound stellar clusters (Kruijssen 2012), i.e.~the cluster formation efficiency (CFE or $\Gamma$, see Bastian 2008). In this model, the CFE ranges from $\Gamma\sim1\%$ in low-pressure galaxies ($P/k<10^4~{\rm K}~{\rm cm}^{-3}$) to $\Gamma\sim50\%$ in high-pressure galaxies ($P/k>10^6~{\rm K}~{\rm cm}^{-3}$), which is quantitatively consistent with recent observations (Goddard et al.~2010; Adamo et al.~2011; Silva-Villa et al.~2013). We thus see that the high-pressure conditions seen in high-redshift galaxies promote the formation of bound stellar clusters.

After determining which fraction of the star formation rate results in bound clusters, this {\it cluster formation rate} must be distributed over some range of cluster masses. Observations (Portegies Zwart et al.~2010) and theory (Elmegreen \& Falgarone 1996) show that the initial cluster mass function (ICMF) follows a power law ${\rm d}N/{\rm d}M\propto M^\alpha$ with index $\alpha=-2$. At the low-mass end, the ICMF continues down to the detection limit. Based on studies of cluster formation in the solar neighbourhood, a commonly-adopted lower mass limit is $M_{\rm min}=100~{\rm M}_\odot$ (Lada \& Lada 2003). At the high-mass end, the ICMF is limited by the Toomre (1964) mass, i.e.~the largest mass scale for gravitational instability in a differentially rotating disc (Hughes et al.~2013; Kruijssen 2014). Because the Toomre mass increases with the ambient gas pressure, the high-pressure conditions seen in high-redshift galaxies enable the formation of clusters much more massive ($M_{\rm max}\sim10^7~{\rm M}_\odot$) than those forming in low-redshift discs ($M_{\rm max}\sim10^5~{\rm M}_\odot$, see e.g.~Larsen 2009).

\subsection{Disruption Phase 1: Early evolution and migration}
The vast majority of stars in the Universe formed within gaseous galaxy discs rather than irregular dwarf galaxies or mergers (Genzel et al.~2010; Rodighiero et al.~2011). We can therefore assume that after their formation, young stellar clusters reside within the gas-rich disc of their host galaxy. Within such discs, the frequent encounters with giant molecular clouds (or analogously with the massive clumps seen in high-redshift galaxies) lead to the rapid disruption of young clusters by impulsive tidal shocks (Spitzer 1987; Gieles et al.~2006). This disruption agent dominates over the more gradual mass loss by evaporation (Kruijssen et al.~2011). in Kruijssen (2015), we show that including this {\it rapid-disruption phase} in the evolutionary history of GCs is crucial for reproducing their present-day properties. If the rapid-disruption phase in the host galaxy disc is excluded, the total dynamical mass loss over a Hubble time is severely underestimated.

The rapid disruption of young GCs in the host galaxy disc comes to an end when the clusters migrate out of the volume that is occupied by the (molecular) gas. In dwarf galaxies, the bursty nature of star formation could lead to changes of the gravitational potential that facilitate this type of migration (Pontzen \& Governato 2012), but across the galaxy mass range the consistently most plausible migration agent seems to be the galaxy mergers associated with hierarchical galaxy formation (White \& Frenk 1991; Kravtsov \& Gnedin 2005), which take place on $\sim{\rm Gyr}$ time-scales (Kruijssen 2015). This type of migration can take two different forms. If the GCs' host galaxy undergoes a major merger, the redistribution of material naturally migrates the young GCs into the halo of the merger remnant (Kruijssen et al.~2012b). If the GCs' host galaxy is cannibalised by a much (i.e.~by a factor of $>3$) more massive galaxy, the tidal stripping of the system also migrates the GCs into the halo of the more massive galaxy. The common denominator is that as long as the host galaxy merges with a galaxy of a similar or larger mass, the rapid-disruption phase comes to an end.

\begin{figure}
\center\includegraphics[width=10.2cm]{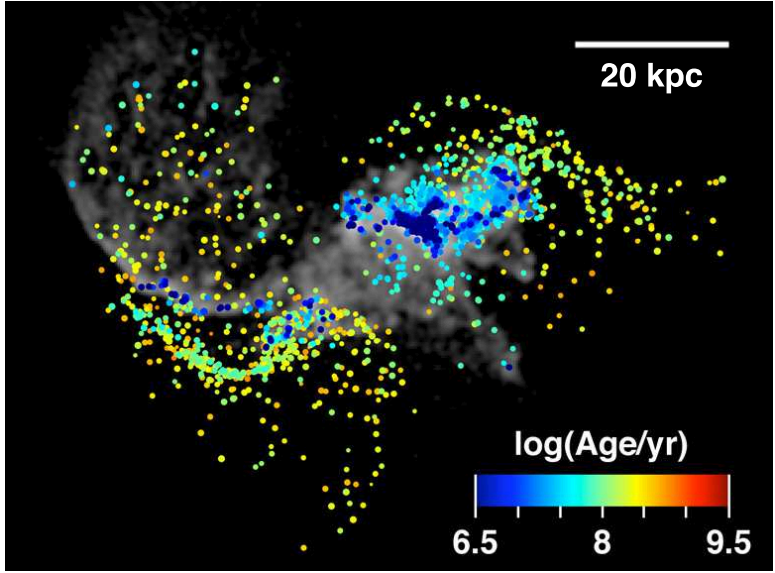}
\caption{Snapshot of galaxy merger model {\tt 1m11} from Kruijssen et al.~(2012b), which includes a sub-grid model for the formation and evolution of the stellar cluster population. Shown is a merger of two Milky Way-mass galaxies at the time of their first encounter. Coloured dots represent stellar clusters, colour-coded by their ages as indicated by the legend. The grey scale indicates the gas surface density. The snapshot shows how intermediate-age clusters are escaping into the halo, whereas those clusters that formed during the merger reside in the gas-rich, disruptive environment of the discs. This cluster migration process is also seen in observations of nearby galaxy mergers (Bastian et al.~2009).}\label{fig:mig}
\end{figure}
In Kruijssen et al.~(2011, 2012b), we presented numerical simulations of galaxy discs and galaxy mergers with a sub-grid model for the formation and evolution of the stellar cluster population. The cluster populations modelled in these simulations (see Figure~\ref{fig:mig}) provide a good match to the observed age distributions of clusters in M83 (Adamo \& Bastian 2015) and in the Antennae galaxies (Kruijssen 2011), and quantitatively show that the survival chances of stellar clusters indeed increase dramatically by migration into the gas-poor galaxy halo during major mergers. 

\subsection{Disruption Phase 2: Quiescent evolution}
Once the (by then intermediate-age) GCs have migrated into the host galaxy halo, their long-term survival has become likely, as their further disruption until the present day is dominated by gradual evaporation. Previous GC formation models have focussed on this phase to model the emergence of the observed GC population, and can be divided in roughly two categories.
\begin{itemize}
\item[(i)]
One (`environmentally independent') family of models accounts for evaporation-driven mass loss using the classical expression by Spitzer (1987), which is independent of the tidal field and is exclusively set by the mass and radius of the GC under consideration (Fall \& Zhang 2001; Prieto \& Gnedin 2008; McLaughlin \& Fall 2008; Li \& Gnedin 2014).
\item[(ii)]
The other (`environmentally dependent') family of models accounts for the results of $N$-body simulations showing that the evaporation-driven mass loss rate of GCs is exclusively set by the tidal field strength (i.e.~the galactic environment) and the GC mass, and is independent of the cluster radius (Vesperini \& Heggie 1997; Baumgardt \& Makino 2003; Gieles \& Baumgardt 2008).
\end{itemize}
Both model families have their own problems. The first family reproduces the near-universal shape of the GC mass spectrum not including an environmental dependence, but is physically inconsistent with $N$-body simulations of evaporating clusters as well as with observations of cluster populations, which show clear indications of environmentally-dependent cluster disruption (e.g.~Bastian et al.~2012). The second family accounts for the environmental variation of the evaporation rate, but as a result cannot explain the near-universality of the GC mass spectrum or the environmental independence of the specific frequency after dividing out the metallicity dependence (Kruijssen 2014).

In the GC formation model described here (Kruijssen 2015), the above problems are alleviated, because most of the GC disruption takes place during their early evolution in the gas-rich host galaxy disc. In this model, the subsequent redistribution of GCs into galaxy haloes erases the correlation between their mass loss history and the present-day environment. Instead, the total mass loss of GCs should correlate with their formation environment, as is corroborated by the observed decrease of the specific frequency with the metallicity at constant galactocentric radius (Harris \& Harris 2002; Lamers et al.~2015).

The relative unimportance of GC evaporation in galaxy haloes to their present-day statistics does not imply the process is uninteresting. For instance, this quiescent phase in the history of GCs is crucial in setting their structural properties (Gieles et al.~2011) as well as their stellar mass functions (Kruijssen 2009). Both areas still provide a broad range of unanswered questions, which are beyond the scope of this paper.

\subsection{Connection to the host galaxy}
All of the discussed quantities governing the formation, disruption, migration, and evaporation of GCs depend on the properties of the natal galaxy. While present-day GCs are no longer associated with the galaxy in which they formed, their metallicities and ages provide clues to their formation environments. The median age of the Galactic GC population ($\tau\sim11.5~{\rm Gyr}$, see Forbes \& Bridges 2010) indicates they typically formed at redshift $z\sim3$. By using the galaxy mass-metallicity relation observed at that redshift (Erb et al.~2006; Mannucci et al.~2009), the metallicities of present-day GCs can be connected to the masses of the galaxies in which they were born.

After identifying the host galaxy mass and metallicity, we adopt an equilibrium-disc model to describe star and cluster formation, as well as the subsequent rapid cluster disruption and migration. As shown by Krumholz \& McKee (2005), such a model is entirely set by the combination of the gas surface density, angular velocity, and Toomre stability parameter (here assumed to be $Q=1$, indicating marginal stability). The gas surface density is set by the ISM pressure (which, as explained above, follows from the maximum GC mass-scale), and the angular velocity is provided by the observed scaling relation between galaxy mass and rotation rate in $z=1.1$--$3.5$ galaxies (F\"{o}rster Schreiber et al.~2009). As a result, we have a working model for star and cluster formation and early evolution as a function of the metallicity of each present-day GC.

In Kruijssen (2015), we show that the ISM pressure (and hence early cluster disruption rate) in the above model weakly {\it increases} with the galaxy mass (and hence metallicity). In hierarchical galaxy formation models (Springel et al.~2005), the rate of mergers with galaxies of at least the GC host galaxy's own mass $M_{\rm host}$ decreases with $M_{\rm host}$. The migration rate into the galaxy halo of young GCs therefore {\it decreases} with the host galaxy mass (and metallicity). We thus see that the rapid-disruption phase (1) leads to more cluster disruption and (2) lasts longer in massive, high-metallicity galaxies.

\section{Model results: the mass spectrum of Galactic globular clusters}
\begin{figure}
\center\includegraphics[width=11.2cm]{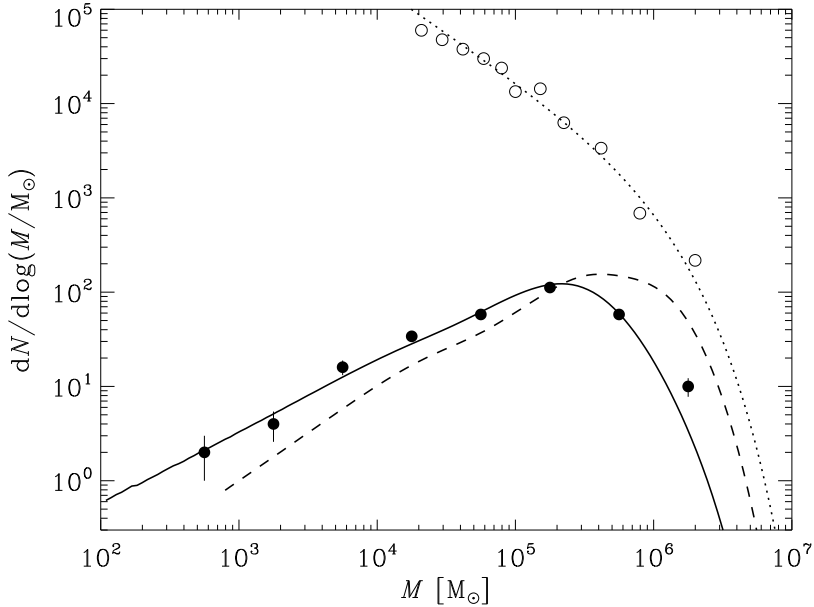}
\caption{Predicted GC mass functions for a Milky Way-like galaxy model that includes both in-situ GC formation and the accretion of GCs from cannibalised dwarf galaxies. The different lines show the GC mass functions at the times of GC formation (dotted) and GC migration into the gas-poor galaxy halo (dashed), as well as at the present day (solid). The open circles indicate the observed mass spectrum of young clusters in the Antennae galaxies (Zhang \& Fall 1999), whereas the solid circles represent the observed mass spectrum of globular clusters in the Milky Way (Harris 1996, 2010 edition). Note that while the model lines reproduce the observational data quite well, they have not been fitted to the observations.} \label{fig:gcmf}
\end{figure}
The above model can be used to predict the GC mass spectrum and specific frequency, as well as the relations of GCs to the host dark matter halo mass and supermassive black hole mass, as a function of GC metallicity across cosmic time. As discussed in Kruijssen (2015), the model simultaneously reproduces these observed distribution functions and scaling relations. In particular, it is the only model to match the decrease of the specific frequency with metallicity as well as its invariance with the galactocentric radius (Harris \& Harris 2002; Lamers et al.~2015), which troubled previous models.

We refer the interested reader to Kruijssen (2015) for an in-depth discussion of the model's formulation and results, as well as its broad range of predictions. Instead of going into a similar level of detail here, we briefly discuss a key result of the model. Figure~\ref{fig:gcmf} shows the predicted GC mass spectrum for a metallicity-composite GC population appropriate for the Milky Way. The different lines show the mass spectra at the times of GC formation ($t=0$) and migration ($t={\rm several~Gyr}$), as well as at the present day ($t=11.5~{\rm Gyr}$). For comparison, it also includes the observed mass spectrum of young clusters in the Antennae galaxies (Zhang \& Fall 1999), as well as the observed mass spectrum of Galactic GCs (Harris 1996, 2010 edition). The figure clearly shows that the adopted initial mass spectrum matches that of young clusters seen forming in the Antennae, and the predicted present-day GC mass spectrum also provides a good match to the observations. We note that the model is not a fit. The good agreement further supports the model's inclusion of the initial rapid-disruption phase in the gas-rich host galaxy disc at the epoch of GC formation.

The discussed model provides a very simple, end-to-end description for GC formation and evolution. It is based on what is currently known of:
\begin{itemize}
\item[(i)]
the formation of stellar clusters in high-pressure conditions such as galaxy mergers (in the local Universe) or clumpy galaxy discs (i.e.~the high-redshift GC formation sites);
\item[(ii)]
cluster disruption by tidal shocks caused by encounters with giant molecular clouds or other substructure in the high-pressure ISM;
\item[(iii)]
cluster migration by galaxy mergers during hierarchical galaxy formation;
\item[(iv)]
gradual cluster disruption by evaporation within galaxy haloes.
\end{itemize}
While the model only represents a first step towards future, more sophisticated end-to-end models, it is the first to combine all relevant phases of GC formation and evolution across cosmic history. In the process, it provides a surprisingly good match to the observed GC population. With this framework in hand, future studies covering the different phases of the model in more detail (e.g.~using numerical simulations) should enable the use of GCs as tracers of high-redshift star formation.

\section*{Acknowledgements}
I am grateful to the conference organisers for the invitation and financial support through an IAU grant, as well as for organising a very lively and stimulating meeting. In addition, I am indebted to my collaborators for their insights and contributions to the projects discussed here. I particularly would like to thank Angela Adamo, Nate Bastian, Michael Hilker, Henny Lamers, Steve Longmore, Jill Rathborne, and Marina Rejkuba.

%\begin{discussion}
%
%\discuss{To be}{inserted by}
%
%\discuss{the}{Editors}
%\end{discussion}


\begin{thebibliography}{60}
\expandafter\ifx\csname natexlab\endcsname\relax\def\natexlab#1{#1}\fi

\bibitem[{{Adamo} \& {Bastian}(2015)}]{adamo15}
{Adamo} A., {Bastian} N., 2015, {The lifecycle of clusters in galaxies}. The
  Birth of Star Clusters, editor S.W. Stahler, Springer, submitted

\bibitem[{{Adamo}, {{\"O}stlin} \& {Zackrisson}(2011){Adamo}, {{\"O}stlin}, \&
  {Zackrisson}}]{adamo11}
{Adamo} A., {{\"O}stlin} G., {Zackrisson} E., 2011, \mnras, 417, 1904

\bibitem[{{Bastian}(2008)}]{bastian08}
{Bastian} N., 2008, \mnras, 390, 759

\bibitem[{{Bastian} {et~al}\mbox{.}(2012){Bastian}, {Adamo}, {Gieles},
  {Silva-Villa}, {Lamers}, {Larsen}, {Smith}, {Konstantopoulos}, \&
  {Zackrisson}}]{bastian12}
{Bastian} N. {et~al.}, 2012, \mnras, 419, 2606

\bibitem[{{Bastian} {et~al}\mbox{.}(2009){Bastian}, {Trancho},
  {Konstantopoulos}, \& {Miller}}]{bastian09}
{Bastian} N., {Trancho} G., {Konstantopoulos} I.~S., {Miller} B.~W., 2009,
  \apj, 701, 607

\bibitem[{{Baumgardt} \& {Makino}(2003)}]{baumgardt03}
{Baumgardt} H., {Makino} J., 2003, \mnras, 340, 227

\bibitem[{{Efremov} \& {Elmegreen}(1998)}]{efremov98}
{Efremov} Y.~N., {Elmegreen} B.~G., 1998, \mnras, 299, 588

\bibitem[{{Elmegreen}(2010)}]{elmegreen10}
{Elmegreen} B.~G., 2010, \apjl, 712, L184

\bibitem[{{Elmegreen} \& {Efremov}(1997)}]{elmegreen97}
{Elmegreen} B.~G., {Efremov} Y.~N., 1997, \apj, 480, 235

\bibitem[{{Elmegreen} \& {Falgarone}(1996)}]{elmegreen96}
{Elmegreen} B.~G., {Falgarone} E., 1996, \apj, 471, 816

\bibitem[{{Erb} {et~al}\mbox{.}(2006){Erb}, {Steidel}, {Shapley}, {Pettini},
  {Reddy}, \& {Adelberger}}]{erb06}
{Erb} D.~K., {Steidel} C.~C., {Shapley} A.~E., {Pettini} M., {Reddy} N.~A.,
  {Adelberger} K.~L., 2006, \apj, 646, 107

\bibitem[{{Fall} \& {Zhang}(2001)}]{fall01}
{Fall} S.~M., {Zhang} Q., 2001, \apj, 561, 751

\bibitem[{{Forbes} \& {Bridges}(2010)}]{forbes10}
{Forbes} D.~A., {Bridges} T., 2010, \mnras, 404, 1203

\bibitem[{{F{\"o}rster Schreiber} {et~al}\mbox{.}(2009){F{\"o}rster Schreiber},
  {Genzel}, {Bouch{\'e}}, {Cresci}, {Davies}, {Buschkamp}, {Shapiro},
  {Tacconi}, {Hicks}, {Genel}, {Shapley}, {Erb}, {Steidel}, {Lutz},
  {Eisenhauer}, {Gillessen}, {Sternberg}, {Renzini}, {Cimatti}, {Daddi},
  {Kurk}, {Lilly}, {Kong}, {Lehnert}, {Nesvadba}, {Verma}, {McCracken},
  {Arimoto}, {Mignoli}, \& {Onodera}}]{forsterschreiber09}
{F{\"o}rster Schreiber} N.~M. {et~al.}, 2009, \apj, 706, 1364

\bibitem[{{Genzel} {et~al}\mbox{.}(2010){Genzel}, {Tacconi}, {Gracia-Carpio},
  {Sternberg}, {Cooper}, {Shapiro}, {Bolatto}, {Bouch{\'e}}, {Bournaud},
  {Burkert}, {Combes}, {Comerford}, {Cox}, {Davis}, {Schreiber},
  {Garcia-Burillo}, {Lutz}, {Naab}, {Neri}, {Omont}, {Shapley}, \&
  {Weiner}}]{genzel10}
{Genzel} R. {et~al.}, 2010, \mnras, 407, 2091

\bibitem[{{Gieles} \& {Baumgardt}(2008)}]{gieles08}
{Gieles} M., {Baumgardt} H., 2008, \mnras, 389, L28

\bibitem[{{Gieles}, {Heggie} \& {Zhao}(2011){Gieles}, {Heggie}, \&
  {Zhao}}]{gieles11b}
{Gieles} M., {Heggie} D.~C., {Zhao} H., 2011, \mnras, 413, 2509

\bibitem[{{Gieles} {et~al}\mbox{.}(2006){Gieles}, {Portegies Zwart},
  {Baumgardt}, {Athanassoula}, {Lamers}, {Sipior}, \& {Leenaarts}}]{gieles06}
{Gieles} M., {Portegies Zwart} S.~F., {Baumgardt} H., {Athanassoula} E.,
  {Lamers} H.~J.~G.~L.~M., {Sipior} M., {Leenaarts} J., 2006, \mnras, 371, 793

\bibitem[{{Goddard}, {Bastian} \& {Kennicutt}(2010){Goddard}, {Bastian}, \&
  {Kennicutt}}]{goddard10}
{Goddard} Q.~E., {Bastian} N., {Kennicutt} R.~C., 2010, \mnras, 405, 857

\bibitem[{{Goodwin} \& {Bastian}(2006)}]{goodwin06}
{Goodwin} S.~P., {Bastian} N., 2006, \mnras, 373, 752

\bibitem[{{Harris}(1996)}]{harris96}
{Harris} W.~E., 1996, \aj, 112, 1487

\bibitem[{{Harris} \& {Harris}(2002)}]{harris02}
{Harris} W.~E., {Harris} G.~L.~H., 2002, \aj, 123, 3108

\bibitem[{{Hills}(1980)}]{hills80}
{Hills} J.~G., 1980, \apj, 235, 986

\bibitem[{{Hodge} {et~al}\mbox{.}(2012){Hodge}, {Carilli}, {Walter}, {de Blok},
  {Riechers}, {Daddi}, \& {Lentati}}]{hodge12}
{Hodge} J.~A., {Carilli} C.~L., {Walter} F., {de Blok} W.~J.~G., {Riechers} D.,
  {Daddi} E., {Lentati} L., 2012, \apj, 760, 11

\bibitem[{{Hughes} {et~al}\mbox{.}(2013){Hughes}, {Meidt}, {Schinnerer},
  {Colombo}, {Pety}, {Leroy}, {Dobbs}, {Garc{\'{\i}}a-Burillo}, {Thompson},
  {Dumas}, {Schuster}, \& {Kramer}}]{hughes13}
{Hughes} A. {et~al.}, 2013, \apj, 779, 44

\bibitem[{{Kravtsov} \& {Gnedin}(2005)}]{kravtsov05}
{Kravtsov} A.~V., {Gnedin} O.~Y., 2005, \apj, 623, 650

\bibitem[{{Kruijssen}(2009)}]{kruijssen09c}
{Kruijssen} J.~M.~D., 2009, \aap, 507, 1409

\bibitem[{{Kruijssen}(2011)}]{kruijssen11c}
{Kruijssen} J.~M.~D., 2011, PhD thesis, Utrecht University, The Netherlands

\bibitem[{{Kruijssen}(2012)}]{kruijssen12d}
{Kruijssen} J.~M.~D., 2012, \mnras, 426, 3008

\bibitem[{{Kruijssen}(2014)}]{kruijssen14c}
{Kruijssen} J.~M.~D., 2014, Classical and Quantum Gravity, 31, 244006

\bibitem[{{Kruijssen}(2015)}]{kruijssen15b}
{Kruijssen} J.~M.~D., 2015, \mnras~in press, arXiv:1509.02163

\bibitem[{{Kruijssen} \& {Longmore}(2013)}]{kruijssen13c}
{Kruijssen} J.~M.~D., {Longmore} S.~N., 2013, \mnras, 435, 2598

\bibitem[{{Kruijssen} {et~al}\mbox{.}(2012{\natexlab{a}}){Kruijssen},
  {Maschberger}, {Moeckel}, {Clarke}, {Bastian}, \& {Bonnell}}]{kruijssen12}
{Kruijssen} J.~M.~D., {Maschberger} T., {Moeckel} N., {Clarke} C.~J., {Bastian}
  N., {Bonnell} I.~A., 2012{\natexlab{a}}, \mnras, 419, 841

\bibitem[{{Kruijssen} {et~al}\mbox{.}(2012{\natexlab{b}}){Kruijssen},
  {Pelupessy}, {Lamers}, {Portegies Zwart}, {Bastian}, \&
  {Icke}}]{kruijssen12c}
{Kruijssen} J.~M.~D., {Pelupessy} F.~I., {Lamers} H.~J.~G.~L.~M., {Portegies
  Zwart} S.~F., {Bastian} N., {Icke} V., 2012{\natexlab{b}}, \mnras, 421, 1927

\bibitem[{{Kruijssen} {et~al}\mbox{.}(2011){Kruijssen}, {Pelupessy}, {Lamers},
  {Portegies Zwart}, \& {Icke}}]{kruijssen11}
{Kruijssen} J.~M.~D., {Pelupessy} F.~I., {Lamers} H.~J.~G.~L.~M., {Portegies
  Zwart} S.~F., {Icke} V., 2011, \mnras, 414, 1339

\bibitem[{{Kruijssen} \& {Portegies Zwart}(2009)}]{kruijssen09b}
{Kruijssen} J.~M.~D., {Portegies Zwart} S.~F., 2009, \apjl, 698, L158

\bibitem[{{Krumholz} \& {McKee}(2005)}]{krumholz05}
{Krumholz} M.~R., {McKee} C.~F., 2005, \apj, 630, 250

\bibitem[{{Lada} \& {Lada}(2003)}]{lada03}
{Lada} C.~J., {Lada} E.~A., 2003, \araa, 41, 57

\bibitem[{{Lada}, {Margulis} \& {Dearborn}(1984){Lada}, {Margulis}, \&
  {Dearborn}}]{lada84}
{Lada} C.~J., {Margulis} M., {Dearborn} D., 1984, \apj, 285, 141

\bibitem[{{Lamers} {et~al}\mbox{.}(2015){Lamers}, {Kruijssen}, {Bastian},
  {Rejkuba}, {Hilker}, \& {Kissler-Patig}}]{lamers15}
{Lamers} H.~J.~G.~L.~M., {Kruijssen} J.~M.~D., {Bastian} N., {Rejkuba} M.,
  {Hilker} M., {Kissler-Patig} M., 2015, in preparation

\bibitem[{{Larsen}(2009)}]{larsen09}
{Larsen} S.~S., 2009, \aap, 494, 539

\bibitem[{{Li} \& {Gnedin}(2014)}]{li14}
{Li} H., {Gnedin} O.~Y., 2014, \apj, 796, 10

\bibitem[{{Longmore} {et~al}\mbox{.}(2014){Longmore}, {Kruijssen}, {Bastian},
  {Bally}, {Rathborne}, {Testi}, {Stolte}, {Dale}, {Bressert}, \&
  {Alves}}]{longmore14}
{Longmore} S.~N. {et~al.}, 2014, Protostars and Planets VI, 291

\bibitem[{{Madau} \& {Dickinson}(2014)}]{madau14}
{Madau} P., {Dickinson} M., 2014, \araa, 52, 415

\bibitem[{{Mannucci} {et~al}\mbox{.}(2009){Mannucci}, {Cresci}, {Maiolino},
  {Marconi}, {Pastorini}, {Pozzetti}, {Gnerucci}, {Risaliti}, {Schneider},
  {Lehnert}, \& {Salvati}}]{mannucci09}
{Mannucci} F. {et~al.}, 2009, \mnras, 398, 1915

\bibitem[{{McLaughlin} \& {Fall}(2008)}]{mclaughlin08}
{McLaughlin} D.~E., {Fall} S.~M., 2008, \apj, 679, 1272

\bibitem[{{Pontzen} \& {Governato}(2012)}]{pontzen12}
{Pontzen} A., {Governato} F., 2012, \mnras, 421, 3464

\bibitem[{{Portegies Zwart}, {McMillan} \& {Gieles}(2010){Portegies Zwart},
  {McMillan}, \& {Gieles}}]{portegieszwart10}
{Portegies Zwart} S.~F., {McMillan} S.~L.~W., {Gieles} M., 2010, \araa, 48, 431

\bibitem[{{Prieto} \& {Gnedin}(2008)}]{prieto08}
{Prieto} J.~L., {Gnedin} O.~Y., 2008, \apj, 689, 919

\bibitem[{{Rathborne} {et~al}\mbox{.}(2015){Rathborne}, {Longmore}, {Jackson},
  {Alves}, {Bally}, {Bastian}, {Contreras}, {Garay}, {Foster}, {Kruijssen},
  {Testi}, \& {Walsh}}]{rathborne15}
{Rathborne} J.~M. {et~al.}, 2015, \apj~in press, arXiv:1501.07368

\bibitem[{{Rodighiero} {et~al}\mbox{.}(2011){Rodighiero}, {Daddi},
  {Baronchelli}, {Cimatti}, {Renzini}, {Aussel}, {Popesso}, {Lutz}, {Andreani},
  {Berta}, {Cava}, {Elbaz}, {Feltre}, {Fontana}, {F{\"o}rster Schreiber},
  {Franceschini}, {Genzel}, {Grazian}, {Gruppioni}, {Ilbert}, {Le Floch},
  {Magdis}, {Magliocchetti}, {Magnelli}, {Maiolino}, {McCracken}, {Nordon},
  {Poglitsch}, {Santini}, {Pozzi}, {Riguccini}, {Tacconi}, {Wuyts}, \&
  {Zamorani}}]{rodighiero11}
{Rodighiero} G. {et~al.}, 2011, \apjl, 739, L40

\bibitem[{{Shapiro}, {Genzel} \& {F{\"o}rster Schreiber}(2010){Shapiro},
  {Genzel}, \& {F{\"o}rster Schreiber}}]{shapiro10}
{Shapiro} K.~L., {Genzel} R., {F{\"o}rster Schreiber} N.~M., 2010, \mnras, 403,
  L36

\bibitem[{{Silva-Villa}, {Adamo} \& {Bastian}(2013){Silva-Villa}, {Adamo}, \&
  {Bastian}}]{silvavilla13}
{Silva-Villa} E., {Adamo} A., {Bastian} N., 2013, \mnras, 436, L69

\bibitem[{{Spitzer}(1987)}]{spitzer87}
{Spitzer} L., 1987, {Dynamical evolution of globular clusters}. Princeton, NJ,
  Princeton University Press, 1987, 191 p.

\bibitem[{{Springel} {et~al}\mbox{.}(2005){Springel}, {White}, {Jenkins},
  {Frenk}, {Yoshida}, {Gao}, {Navarro}, {Thacker}, {Croton}, {Helly},
  {Peacock}, {Cole}, {Thomas}, {Couchman}, {Evrard}, {Colberg}, \&
  {Pearce}}]{springel05d}
{Springel} V. {et~al.}, 2005, \nat, 435, 629

\bibitem[{{Swinbank} {et~al}\mbox{.}(2011){Swinbank}, {Papadopoulos}, {Cox},
  {Krips}, {Ivison}, {Smail}, {Thomson}, {Neri}, {Richard}, \&
  {Ebeling}}]{swinbank11}
{Swinbank} A.~M. {et~al.}, 2011, \apj, 742, 11

\bibitem[{{Tacconi} {et~al}\mbox{.}(2013){Tacconi}, {Neri}, {Genzel}, {Combes},
  {Bolatto}, {Cooper}, {Wuyts}, {Bournaud}, {Burkert}, {Comerford}, {Cox},
  {Davis}, {F{\"o}rster Schreiber}, {Garc{\'{\i}}a-Burillo}, {Gracia-Carpio},
  {Lutz}, {Naab}, {Newman}, {Omont}, {Saintonge}, {Shapiro Griffin}, {Shapley},
  {Sternberg}, \& {Weiner}}]{tacconi13}
{Tacconi} L.~J. {et~al.}, 2013, \apj, 768, 74

\bibitem[{{Toomre}(1964)}]{toomre64}
{Toomre} A., 1964, \apj, 139, 1217

\bibitem[{{Vesperini} \& {Heggie}(1997)}]{vesperini97b}
{Vesperini} E., {Heggie} D.~C., 1997, \mnras, 289, 898

\bibitem[{{White} \& {Frenk}(1991)}]{white91}
{White} S.~D.~M., {Frenk} C.~S., 1991, \apj, 379, 52

\bibitem[{{Zhang} \& {Fall}(1999)}]{zhang99}
{Zhang} Q., {Fall} S.~M., 1999, \apjl, 527, L81

\end{thebibliography}
\end{document}